\documentstyle[12pt]{article}

\def\hb{\hfill\break}
\def\ih{\'\i}
\def\hb{\hfill\break}

\title{THE BATALIN-TYUTIN  FORMALISM ON THE COLLECTIVE COORDINATES QUANTISATION
OF THE SU(2) SKYRME MODEL}

\author{Wilson Oliveira\thanks{e-mail:wilson@fisica.ufjf.br}
 and Jorge Ananias Neto\thanks{e-mail:jorge@fisica.ufjf.br}\\
 Departamento de F\ih sica, ICE \\ Universidade Federal de Juiz de
Fora, 36036-330 \\ Juiz de Fora, MG, Brazil }

\date{ }

\begin{document}

\maketitle

\begin{abstract}
\noindent We apply The Batalin-Tyutin constraint formalism  of
converting a second class system into a first class system for
the rotational quantisation of the SU(2) Skyrme model. We
obtain the first class constraint and the Hamiltonian in the
extended phase space. The vacuum functional is constructed
and evaluated in the unitary gauge and a multiplier dependent
gauge. Finally, we discuss the spectrum of the extended
theory. The use of the BT formalism on the collective
coordinates quantisation of the SU(2) Skyrme model leads
an additional term in the usual quantum Hamiltonian that
can improve the phenomenology predicted by the Skyrme model.

\end{abstract}

\vskip 1 cm


\newpage

\section{Introduction}

Most of the static properties of the Nucleon and the Delta are well
reproduced by the Skyrme model\cite{Skyrme}, which describes baryons
and their interactions
through soliton solutions in a nonlinear type sigma model.
If  our goal is to improve the usual
phenomenology predicted by the Skyrme model then, we believe that there are
two simple ways
to deal: first, in the classical sector, we can introduce higher order
term in the derivatives of
the pion field\cite{Marleau}. This procedure leads the physical
values to approach to the correct experimental values; second, in the
quantum sector, we can analyze with more detail the process of
the collective coordinates quantisation of the Skyrme model\cite{ANW}. Thus,
the purpose of this
paper is to develop the latter option, investigating with more rigorous
the SU(2) Skyrme model in quantum mechanics. Some authors\cite{Fujii} have
pointed out that when the quantisation is performed with more care,
an extra term appears in the usual Skyrme quantum Hamiltonian. The aim
of the present work is to employ the BT\footnote{ The Batalin Tyutin
formalism is a powerful
framework for the quantisation of gauge field theories
that can be used when we depart with nontrivial Dirac
brackets relations.} constraint formalism
\cite{BT}with the objective to overcome the problem of operator ordering that
occurs in the Dirac brackets of the collective coordinates operators
\cite{J A Neto}
\par The remaining part of this paper is organized as follows. In
sec 2 we give a short outline of the BT formalism. In sec. 3 we
apply the BT method in the SU(2) Skyrme model when we derive the
first class constraints, an involutive extended Hamiltonian and the
vacuum functional Z for two different gauge fixing. In
sec 4  by imposing strongly the first class constraint, we obtain
the usual mass spectrum with an additional term. Finally, in
sec 5 we give the conclusions.

\section {Brief review of the BT formalism}

\noindent Let us consider a system described by a Hamiltonian $H_0$
in a phase space $(\phi^i(x), \pi_i(x))$ with i=1,...,N. We assume
that the fields are bosonic( extension to include fermionic degrees
of freedom and to the continuous case can be done in a
straightforward way). We also suppose that this system contains a
set of linearly independent bosonic second class constraints.
Denoting  $T_\alpha=T_\alpha(\phi,\pi)$, with $\alpha =1,...,M<2N$, then
the matrix

\begin{equation}
\label{pb}
\Delta_{\alpha\beta} (x,y)=\{ T_\alpha (x),T_\beta (y)\},
\end{equation}

\noindent has a nonvanishing determinant. The inclusion of other
constraints (i.e. fermion or first class) is a matter of technical
detail and poses no problems in developing the formalism.
\par The general philosophy of the BT formalism is to convert
second class constraint into first class ones. This is achieved
by introducing new dynamical fields, one for each second class
constraint. We note that the connection between the number of
second class constraint and the new fields in one to one is to
keep the same number of the physical degrees of freedom in the
resulting extended theory. We denote these new fields by
$\Psi^a(x)$ and assume that they have the following Poisson
algebra

\begin{equation}
\label{pba}
\{ \Psi^\alpha(x),\Psi^\beta(y) \}=\omega^{\alpha\beta}(x,y),
\end{equation}

\noindent where $\omega^{\alpha\beta}$ is an invertible field independent
antisymmetric matrix. The new dynamical fields are introduced
to extend the original phase space

\begin{equation}
\label{ps}
(\phi,\pi) \oplus (\Psi).
\end{equation}

\noindent The new first class constraints of the system in the extended phase
space (\ref{ps}) are denoted by $\tilde{T_\alpha}$ . Of course, these
depend on the new fields $\Psi^\alpha$, written as

\begin{equation}
\label{tal}
\tilde{T_\alpha}=\tilde{T_\alpha}(\phi,\pi,\Psi),
\end{equation}

\noindent and satisfy the boundary condition

\begin{equation}
\label{tal1}
\tilde{T_\alpha}(\phi,\pi,0)=T_\alpha(\phi,\pi),
\end{equation}

\noindent where the right hand side of (\ref{tal1}) is just the original
set of second class constraint. The characteristic of the new
constraints is that they are assumed to be strongly involutive, i.e.

\begin{equation}
\label{inv}
\{\tilde{T_\alpha},\tilde{T_\beta}\}=0.
\end{equation}

\noindent The solution of (\ref{inv}) can be achieved
by considering $\tilde{T_\alpha}$ expanded as a power series
expansion is

\begin{equation}
\label{series}
\tilde{T_\alpha}=\sum^\infty_{n=0} T^{(n)}_\alpha,
\end{equation}

\noindent

\noindent where $T^{(n)}_\alpha$ is a term of order n in $\Psi$.
Compatibility with the boundary condition (\ref{tal1}) requires
that $T^{(0)}_\alpha=T_\alpha$. The first order correction term
in the infinite series\cite{BT} is

\begin{equation}
\label{T1}
T^{(1)}_\alpha(x) = \int dy X_{\alpha\beta}(x,y) \Psi^\beta(y),
\end{equation}

\noindent and the involutive bracket(\ref{inv}) leads the follow
relation

\begin{equation}
\label{xab}
\int dz dz' X_{\alpha\mu}(x,z) \omega^{\mu\nu}(z,z')
 X_{\nu\beta}(z',y) = -\Delta_{\alpha\beta}(x,y) .
\end{equation}

\noindent This determines $T^{(1)}_\alpha$. Equation (\ref{xab})
does not give $X_{\alpha\mu}$ univocally, because it also contains
the still unknown $ \omega^{\mu\nu}$. What we usually do is to
choose $\omega^{ab}$ in such a way that the new fields are
unconstrained. It is opportune
to mention that this procedure is not always possible to be done
\cite{Barcelos}. It is possible to show\cite{BT} that when
$X_{\alpha\beta}$ does not depend on $\phi$ and $\pi$, only
$T_\alpha^{(1)}$ contributes in the series(\ref{series}), defining the
first class constraint.
\par The next step in the BT formalism is that any dynamical function
$H(\phi,\pi)$ (in instance, the Hamiltonian) has also to be properly
modified in order to be strongly involutive with the first class
constraints $\tilde{T}_\alpha$. Denoting the strongly involutive
function by $\tilde{H}$, then we have

\begin{equation}
\label{Hinv}
\{\tilde{H},\tilde{T_\alpha}\}=0,
\end{equation}

\noindent subject to the boundary condition $\tilde{H}(\phi,\pi,0)=
H_c(\phi,\pi)$, where $H_c$ is the canonical Hamiltonian. The general
solution\cite{BT} for the involutive
Hamiltonian $\tilde{H}$, which can be expanded in an infinite series,
is

\begin{equation}
\label{Hcorr}
\tilde{H}=H_c+\sum_{n=1}^\infty H^{( n)},
\end{equation}

\noindent where $H^{( n)}$ is given by
\begin{equation}
\label{gsol}
H^{(n)} ~=~ -{1\over n} \int dx dy dz \Phi^i(x)
\omega_{ij}(x,y) X^{jk}(y,z) G_k^{(n-1)}(z), \, (n\geq1),
\end{equation}

\noindent and the generating functions $G_k^{(n)}$,
in the case that only $T_\alpha^{(1)}$
contributes in the series(\ref{series}), read

\begin{eqnarray}
\label{gene}
G_i^{(0)} = \{T_i^{(0)}, H_c \}_{(\phi\pi)}, \nonumber \\
G_i^{(n)} = \{T_i^{(0)}, H^{(n)}\}_{(\phi\pi)}
+ \{T_i^{(1)}, H^{(n-1)} \}_{(\phi\pi)}, \,\, (n\geq1).
\end{eqnarray}

\noindent Here, $\omega_{ij}$ and $X^{ij}$ are the inverse
matrices of $\omega^{ij}$ and $X_{ij}$ respectively. This
concludes this brief review on the BT construction of the
first class system which is strongly involutive.

\section {The BT formalism on the SU(2) Skyrme model}

\noindent The Lagrangian of the SU(2) Skyrme model performed
in a semi-classical collective coordinates expansion\cite{ANW}
reads

\begin{equation}
\label{Lag}
L = - M + \lambda Tr [ \partial_0 A\partial_0 A^{-1} ]
= -M + 2 \lambda \dot{a}^i \dot{a}^i,
\end{equation}

\noindent where M is the soliton mass, $\lambda$ is the inertia
moment and A is a SU(2) matrix which can be expanded as
$A=a^0+a.\tau$. The primary constraint is

\begin{equation}
\label{pri}
T_1 = a^ia^i - 1 \approx 0.
\end{equation}

\noindent Introducing the conjugate momentum

\begin{equation}
\label{cm}
\pi^i = {\partial L \over \partial \dot{a}_i} = 4 \lambda \dot{a}^i,
\end{equation}

\noindent we can now rewrite the canonical Hamiltonian in the form

\begin{equation}
\label{chr}
H_c=\pi^i \dot a^i-L=4\lambda \dot a^i \dot a^i -L=M+2 \lambda \dot a^i
\dot a^i
=M+{1\over 8 \lambda } \sum_i {\pi^i}^2.
\end{equation}

\noindent Then, the standard quantization is made where we replace
$\,\pi^i \,$ by $\, -i \partial/\partial a_i \,$ in (\ref{chr}),
 leading to

\begin{equation}
\label{uqh}
H=M+{1\over 8 \lambda } \sum_{i=0}^3 (-{\partial
\over\partial{a_i}^2}).
\end{equation}

\noindent A typical polynomial wavefunction
$(a^0 + i a^1)^l$ is an eigenvector of the Hamiltonian
(\ref{uqh}), with the eigenvalues given by \footnote
{This wave function is also eigenvector of the spin and
isospin operators, written as\cite{ANW} $ J^k={1\over 2}
( a_0 \pi_k -a_k \pi_0 - \epsilon_{klm} a_l \pi_m )$  and
$ I^k={1\over 2 } ( a_k \pi_0 -a_0 \pi_k- \epsilon_{klm} a_l
\pi_m ).$}.

\begin{equation}
\label{uqhe}
E=M+{1\over 8 \lambda } l(l+2), \,\,\,\, l=1,2,\dots \,\,.
\end{equation}

\noindent Following the usual Dirac standard procedure\cite{Dirac}
we find a secondary constraint,

\begin{equation}
\label{sc}
T_2 = a^i\pi_i \approx 0
\end{equation}

\noindent obtained by conserving $T_1$ with the total Hamiltonian

\begin{equation}
\label{tH}
H_T = H_c + \lambda_c T_1,
\end{equation}

\noindent where $\lambda_c$ is a Lagrange multiplier. No further
constraints are generated via this iterative procedure.
The constraints $T_1$ and $T_2$ are second-class, satisfying the
Poisson algebra

\begin{equation}
\label{Pa}
\Delta_{\alpha, \beta} = \{T_\alpha,T_\beta\} = -2 \epsilon_{\alpha \beta}
a^ia^i, \,\, \alpha,\beta = 1,2
\end{equation}

\noindent where $\epsilon_{\alpha \beta}$ is the antisymmetric
tensor normalized as $\epsilon_{12} = -\epsilon^{12} = -1.$
\par In order to convert this system into first-class one, the first step
is to transform $T_\alpha$ into the first-class by extending the phase
space. Following the BT formalism\cite{BT}, we introduce new auxiliary
coordinates\footnote{We now rewrite the auxiliary fields defined in
(\ref{pba}) as $b^\alpha$.} $b^i$ to convert the second-class constraint
$T_\alpha$ into the first-class one in the extended phase space,
and consider that the Poisson algebra of these new coordinates
is given by

\begin{equation}
\label{Pan}
\{b^\alpha,b^\beta\} = \omega^{\alpha\beta},
\end{equation}

\noindent where $\omega^{ij}$ is an antisymmetric matrix.
Then, the modified constraint in the extended phase space
is given by

\begin{equation}
\label{mc}
\tilde{T}_\alpha(a^i\pi_i,b^j) = T_\alpha + \sum_{n=1}^\infty T_\alpha^{(n)};
\,\,T_\alpha^{(n)}\sim (b^j)^n,
\end{equation}

\noindent satisfying the boundary condition

\begin{equation}
\label{bonc}
\tilde{T}_\alpha(a^i\pi_i,0) = T_\alpha .
\end{equation}

\noindent To obtain $\tilde{T}_\alpha$ we follow the procedure
discussed in sec. 3. The first order correction term in the infinite
series\cite{BT} is given by

\begin{equation}
\label{for}
T_\alpha^{(1)} = X_{\alpha\beta} b^\beta,
\end{equation}

\noindent and the first-class constraint algebra of $\tilde{T}_\alpha$
requires the condition as follows

\begin{equation}
\label{fcc}
X_{\alpha\mu} \omega^{\mu\nu}X_{\beta\nu} = -\Delta_{\alpha\beta}.
\end{equation}

\noindent A possible choice for $\omega^{\mu\nu}$ and $X_{\alpha\beta}$
satisfying (\ref{Pan}) and (\ref{fcc}) is\cite{BT}

\begin{eqnarray}
\label{apc1}
\omega^{\mu\nu} = 2 \epsilon^{\mu\nu}, \\\
\label{apc2}
X_{\alpha \beta} = \left( \begin{array}{clcr} 1  & \,\,0 \\ 0
& a^i a^i \end{array} \right) \,\,.
\end{eqnarray}

\noindent As was enphasized in Ref.\cite{BT}, there is a natural
arbitrariness in this choice, which corresponds to the canonical
transformation in the extented phase space. Using (\ref{mc}),
(\ref{bonc}), (\ref{for}), (\ref{apc1}) and (\ref{apc2}), the new set of
constraints is found to be

\begin{eqnarray}
\label{nsc}
\tilde{T}_1 = T_1 + b^1, \nonumber \\
\tilde{T}_2 = T_2 - a^ia^i b^2,
\end{eqnarray}

\noindent which are strongly involutive,

\begin{equation}
\label{nsc1}
\{\tilde{T}_\alpha, \tilde{T}_\beta\} = 0.
\end{equation}

\noindent Thus, we have all first-class constraints in the extended
phase space by applying the BT formalism systematically. We observe
further that the terms in the series (\ref{mc}) for $n>1$ are
redundant. This completes the conversion of the second-class
constraints $T_\alpha$ to first-class ones $\tilde{T}_\alpha$.
\par Next, we derive the corresponding involutive Hamiltonian
in the extended phase space. Considering what we have seen in
sec. 3 and that we also have $T^{(n)}=0$ for $n>1$, the corrections
that give $\tilde{H}$ can be written as

\begin{equation}
\label{hcor}
\tilde{H}^{(n)} = - {1 \over n} b^\mu \omega_{\mu\nu}X^{\nu\rho}
 G_\rho^{(n-1)}, \,\, (n \geq 1),
\end{equation}

\noindent where $G_\rho^{(n)}$ is given by (\ref{gene}) and

\begin{eqnarray}
\label{grnw1}
\omega_{\mu\nu} = - {1 \over 2} \,\, \epsilon_{\mu\nu}, \\
\label{grnw2}
X^{\nu\rho} = \left( \begin{array}{clcr} 1 & \,\,\,\,\,0 \\ 0 &
 -{1\over a^ia^i} \end{array} \right).
\end{eqnarray}

\noindent Using the expression for the second-class
constraints $T_\alpha$ (eqs. (\ref{pri}) and (\ref{sc})) and
the Hamiltonian (\ref{chr}) as well as (\ref{hcor}),(\ref{grnw1})
and (\ref{grnw2}), it is possible to compute the terms appearing
in the power series of the involutive Hamiltonian

\begin{eqnarray}
\label{hpower}
\tilde{H} = M + {1\over 8\lambda} [1 - {b^1\over a^ia^i}
+ ({b^1\over a^ia^i})^2 - ({b^1\over a^ia^i})^3
+ \ldots ] \pi_j\pi_j \nonumber \\ -{1\over 4\lambda} b^2
[1 - {b^1\over a^ia^i} +
({b^1\over a^ia^i})^2 - ({b^1\over a^ia^i})^3
+ \ldots ] a^i\pi_j \nonumber \\
+{1\over 8\lambda} (b^2)^2 [1 - {b^1\over a^ia^i}
+ ({b^1\over a^ia^i})^2 - ({b^1\over a^ia^i})^3
+ \ldots ] a^ia^i\,.
\end{eqnarray}

\noindent If the convergence ratio, $\,R=|{b^1
\over a^ia^i}|\,$, is less than one, the extended
canonical Hamiltonian can be summed in a geometric series

\begin{eqnarray}
\label{ech}
\tilde{H} = M + {1 \over 8 \lambda}  { (a^ia^i) \over
a^ia^i + b^1} \pi_j\pi_j- {1 \over 4 \lambda} {(a^ia^i) b^2
\over a^ia^i + b^1}
a^j \pi_j + {1 \over 8 \lambda} { (a^ia^i)^2 (b^2)^2\over
a^ia^i + b^1},
\end{eqnarray}

\noindent which is involutive with the first-class constraints,

\begin{equation}
\label{ifcc}
\{ \tilde{T_\alpha},\tilde{H} \} = 0 \,\,\,\, (\alpha=1,2).
\end{equation}

We now look for the vacuum functional Z. A consistent way of doing
this is by means of the path integral formalism in
Faddev-Senjanovic \cite{FS}. Let us identify the new variables
$b^\mu$ as a canonically conjugate pair $ (\phi, \pi_\phi)$
in the Hamiltonian formalism,

\begin{eqnarray}
\label{cpair}
b^1 \rightarrow 2 \phi \,, \nonumber \\
b^2 \rightarrow \pi_\phi \,,
\end{eqnarray}

\noindent satisfying (\ref{Pan}), (\ref{apc1}) and (\ref{apc2}). Then,
the general expression for the vacuum functional reads

\begin{equation}
\label{gvf}
Z = N \int [d\mu] \exp \{ i \int dt (\dot{a}^i\pi_i +
\dot{\phi}\pi_\phi - \tilde{H}) \,\,,
\end{equation}

\noindent with the measure $[d\mu]$ given by

\begin{equation}
\label{mesure}
[d\mu] = [da^i] [d\pi_i] [d\phi] [d\pi_\phi] \prod_{\alpha,
\beta=1}^2 \delta(\tilde{T}_\alpha) \delta(\chi_\beta)
 | det \{ \tilde{T}_\alpha,\chi_\beta \} | \,\,,
\end{equation}

\noindent where the Hamiltonian is now expressed in terms of
$(\phi,\pi_\phi)$ instead of $b^\mu$. The gauge fixing conditions
$ \chi_\beta$ are chosen so that the determinant occurring in
the functional measure is nonvanishing. let us now compute Z in
different gauges. First, we consider the unitary gauge

\begin{eqnarray}
\label{ugauge}
\chi_1 = a^ia^i - 1, \nonumber \\
\chi_2 = a^i\pi_i.
\end{eqnarray}

\noindent The vacuum functional Z takes the form

\begin{eqnarray}
\label{vfz}
Z = N \int [da^i] [d\pi_i] [d\phi] [d\pi_\phi]\delta(a^ia^i - 1
+2\phi) \nonumber \\  \delta(a^i\pi_i - a^ia^i\pi_\phi)
\delta(a^ia^i-1) \delta(a^i\pi_i) \,\, \exp \{i \int dt (
\dot{a}^i\pi_i + \dot{\phi}\pi_\phi - M - \nonumber \\{1\over 8\lambda}
{a^ia^i \over a^ia^i + 2\phi} \pi_j\pi_j  + {1\over 4\lambda}
{a^ia^i \over a^ia^i + 2\phi} \pi_\phi a^j\pi_j
-{1\over 8\lambda} {(a^ia^i)^2 \over a^ia^i + 2\phi}{\pi_\phi}^2 )\}\,\,.
\end{eqnarray}

\noindent We note that due $\delta(a^ia^i - 1)$ in (\ref{vfz})
the Faddev-Senjanovic determinant can be absorbed in the
normalization. The $\phi$ and $\pi_\phi$ integrations are trivially
performed. After exponentiating the delta function
$\delta(a^i\pi_i)$ with Fourier variable $\xi$, we obtain

\begin{eqnarray}
\label{vfzxi}
Z = N \int [da^i] [d\pi_i] [d\xi] \,\delta(a^ia^i - 1)
\nonumber \\  \exp \{ i \int dt ( \dot{a}^i\pi_i - M -
{1\over 8\lambda} \pi_i\pi_i - \xi a^i\pi_i ) \}.
\end{eqnarray}

\noindent Performing the integration over $\pi_i$, we
finally obtain

\begin{equation}
\label{vfzxif}
Z = N \int [da^i] \, \delta (a^ia^i -1) \,
\exp \{ i \int dt (- M + 2 \lambda
\dot{a}^i\dot{a}^i)\},
\end{equation}

\noindent where we have absorbed the integration over $\xi$
into the normalization. Expression (\ref{vfzxif}) is,
therefore, seen to reproduce the original Lagrangian
(\ref{Lag}) subject to the constraint (\ref{pri}). This result
shows, without doubt, the consistency of the theory . Now, let
us consider a multiplier dependent gauge\cite{BFV}

\begin{eqnarray}
\label{mdg}
\chi_1 = \pi_\phi + p_1 \,, \nonumber \\
\chi_2 = \phi.
\end{eqnarray}

\noindent The vacuum functional Z takes the form

\begin{eqnarray}
\label{vfzmdg}
Z = N \int [da^i] [d\pi_i] [d\phi] [d\pi_\phi]
 \,\delta(a^ia^i - 1+2\phi) \delta(a^i\pi_i-a^ia^i\pi_\phi)
\nonumber \\ \delta(\pi_\phi+p_1) \delta(\phi) |2 a^ia^i|
 \exp \{ i \int dt ( \dot{a}^i\pi_i + \dot{\phi}\pi_\phi
+\dot{\lambda}^1 p_1- M \nonumber \\ -{1\over 8\lambda} {a^ia^i \over
a^ia^i+2\phi} \pi_j\pi_j + {1 \over 4\lambda}
{a^ia^i \over a^ia^i+2\phi} \pi_\phi a^j\pi_j \nonumber \\
- {1 \over 8\lambda} {(a^ia^i)^2 \over a^ia^i+2\phi}
{\pi_\phi}^2)\}.
\end{eqnarray}

\noindent The $\phi$ and $p_1$ integrations are trivially
done. Then, the Faddeev-Senjanovic determinant can be
absorbed in the normalization. Exponentiating the delta
function $\delta(a^i\pi_i - a^ia^i \pi_\phi)$ with Fourier
variable $\xi$, and doing successively the $\pi_i$ and
$\pi_\phi$ integrations, we obtain

\begin{eqnarray}
\label{vfzmdg1}
Z = N \int [da^i] [d\lambda^1] \, \delta (a^ia^i -1) \nonumber \\
\exp \{ i \int dt [- M + 2 \lambda \dot{a}^i\dot{a}^i
- {1 \over 32 \lambda} - (4\lambda\dot{\lambda}^1 + 1)
{\dot{\lambda}^1 \over 2}] \},
\end{eqnarray}

\noindent where we have absorbed the $\xi$ integration into
the normalization. We observe that this expression differs
from  (\ref{vfzxif}), but both are expected to yield
identical S-matrix elements by the Fradkin-Vilkovisky theorem
\cite{FV}. The expression (\ref{vfzmdg1}) illustrates the
generality of the BT-Formalism, since it can not be obtained
by conventional quantisation methods.\cite{FS}

\section{The spectrum of the extended theory}

In order to obtain the spectrum of the extended theory, the
constraints (\ref{nsc}) are treated as strong equations.
Consequently, we can replace the canonical conjugate
pair $(b^1,b^2)$ by the collective coordinates
$(a_i,\pi_i)$ in the Hamiltonian (\ref{ech}), which
reads

\begin{equation}
\label{sequa}
\tilde{H} = M + {1\over 8\lambda} a^ia^i \pi_j\pi_j
-{1\over 4\lambda} a^i\pi_i a^j\pi_j .
\end{equation}

\noindent Now $\pi_j$ describes a free momentum particle
and its representation on the collective coordinates space $a^i$
is given by

\begin{equation}
\label{piconfig}
\pi_j = -i {\partial\over \partial x^j}.
\end{equation}

\noindent If we substitute the expression (\ref{piconfig}) in
(\ref{sequa}), we obtain

\begin{equation}
\label{hquanc}
\tilde{H} = M - {1\over 8\lambda} a^ia^i \partial_j\partial_j
+{1\over 4\lambda} a^j\partial_j + {1\over 4\lambda} a^ia^j
\partial_i\partial_j,
\end{equation}
\begin{sloppypar}

\noindent with the eigenvalues\footnote{Remember that the eigenvectors
 are written in a polynomial form $(a^0+i a^1)^l$.} given by
\end{sloppypar}

\begin{equation}
\label{hquan}
E = M + {1\over 8\lambda} \left\lbrack
 l(l+2) + l(l-2) \right\rbrack .
\end{equation}

\noindent Comparing  expression (\ref{hquan}) with (\ref{uqhe})
 we see that an extra term appears in the last equation.
We can observe that for the Nucleon state, l=1,
the extra term is negative and according to A. Toda\cite{Toda}, this
result improves the physical values given by the Skyrme model.

\section{Conclusions}

In this work we have applied the Batalin-Tyutin formalism
to the rotational quantisation of
the SU(2) Skyrme model. We note that the conventional Dirac
method\cite{Dirac} presents operator ordering difficult
and BT formalism, in principle, overcomes these problems.
We have obtained the involutive Hamiltonian, and when we impose
strongly the news involutive constraints, we get the usual
mass spectrum of the theory plus
an additional term. This result has been found by many
authors\cite{Fujii} using different procedures.
Thus, as we have remarked in the introduction, the
combined results of introducing higher derivative terms
\cite{Marleau},\cite{EU}, and an adequate quantisation which
contains the constrains information,
can lead the Skyrme model to predict,
with more success, the physical parameters of the mesons and
baryons. In a future paper we intend to apply the Batalin-
Tyutin formalism on the quantisation of Skyrme model with
the inclusion of higher derivative terms.

\section{Acknowledgments}

We would like to thank C.S.M. Mendon\c ca for critical reading,
J. Barcelos-Neto for useful discussions, and the Departamento de
Campos e Part\'{\i}culas of Centro Brasileiro de Pesquisas
F\ih sicas for hospitality while part of this work was carried out.

\end{document}